-------------------------------------------------------------------------------
\magnification=1200
\baselineskip=20pt

\line{\hfil MRIPHY/970719}

\def\lam{\Lambda}
\def\mh{m_H}
\def\aefb{A^e_{fb}}
\def\alr{A_{LR}}
\def\sqs{\sqrt{s}}
\def\gmmuu{\gamma^{\mu}}
\def\dmuphi{D_{\mu}\Phi}
\def\dmu{D^{\mu}}
\def\taua{\tau_a}
\def\delmu{\partial_{\mu}}
\def\alp{\alpha}
\def\dalr{\delta A_{LR}}
\def\daefb{\delta A^e_{fb}}
\def\gm{\gamma}
\def\gel{g^e_L}
\def\ger{g^e_R}

\centerline{\bf Effect of contact interactions on higgs production}
\centerline{\bf cross-section at an $e^+e^-$ collider}  
\vskip .4truein
\centerline{\bf Uma Mahanta}
\centerline{\bf Mehta Research Institute}
\centerline{\bf Chhatnag Road, Jhusi}
\centerline{\bf Allahabad-221506, India}
\vskip 1truein
\centerline{\bf Abstract}

New interactions appearing at a scale $\Lambda$ larger than the weak
interaction scale v can affect physical processes at energies below
$\Lambda$ through non-renormalizable $SU(3)\times SU(2)\times U(1)_y$
invariant operators added to the standard model Lagrangian. In this
article we investigate the effect of flavor conserving contact interactions
on the total cross-section for the process $e^+e^-\rightarrow HZ$
 at $\sqs =500$ Gev. We find that for $\Lambda\approx 2.5 $ Tev, which
is consistent with LEP and SLD asymmetry measurements on Z peak as well
as theoretical estimates, these
operators can increase the total cross-section by a factor of 3 relative to
the SM for intermediate mass higgs boson.

\vfill\eject

The SM has been extremely successful in explaining all the experimental
data so far. However in spite of its extraordinary phenomenological success
many theorists regard the SM as an effective low energy theory valid below 
some cut off scale $\lam $ of the order of a few Tev. One of the
 fundamental reasons behind this view
is that the Higgs mass in the SM receives radiative
corrections that diverges quadratically with the cut off $\Lambda$ [1].
Hence to stabilize the higgs mass around the weak scale, which is
the natural upper bound for $\mh $, the cut off scale $\lam $ should be
of the order of a few Tev. New interactions can appear at or above the
 scale $\lam $ 
involving new heavy particles. Their effects on physical processes
at energies below $\lam $ can be described by an effective Lagrangain
 containing $SU(3)_c\times SU(2)\times U(1)_y$ invariant 
non-renormalizable operators involving only the light SM fields [2].
Since we shall be considering the effects of Tev scale
 new physics on higgs
production cross-section the light SM fields should include $\phi $.
For the same reason the gauge symmetry will be assumed to be linearly
realized on the SM fields. 
In 
addition to gauge  symmetries one might also impose other constraints
like baryon number and lepton number
conservation. The non-renormalizable operators
 can be expressed as a systematic power series expansion in ${1\over \lam }$.
The lowest dimensional operator will clearly have the most dominant effect at
energies below $\lam$. Further the effect of these non-renormalizable
 operators increases as the characteristic energy scale of the process
under study approaches $\lam $.

In this article we shall consider the effect of flavor conserving d=6 operators
involving leptons, scalar and gauge fields on intermediate mass ($m_H
\approx $ 100-250 Gev) higss production via the process $e^+e^-\rightarrow
HZ$ at $\sqrt {s}=500$ Gev. We find that for $\lam \approx $ 2.5 Tev the
 flavor conserving operators satisfy the
LEP and SLD constraints on $\aefb $
and $\alr $. However the same value of $\lam$ can increase the cross-section
(if the new physics contribution interferes constructively with that of
SM) for the process by a factor of 3 relative to that of the SM for a higgs
mass of 150-200 Gev. On the other hand for destructive interference
the cross-section decreases by a factor of .7 relative to SM for the same
range of values of $\lam$ and $\mh$.
The production and detection of intermediate higgs will be one of the
important tasks of future $e^+e^-$ collider.
The detection of such a  higgs boson will be extremely difficult
at hadron collider since the higgs decays mainly into $b\bar{b}$
 which can remain hidden in the background arising from $t\bar {t} $
 pair production [3]. It is therefore extremely
 important to consider the effects
of Tev scale new physics on higgs production cross-section at future
$e^+e^-$ colliders. 

The effective Lagrangian is given by $L= L_{SM}+L_{\lam}$ where $L_{\lam}
=\sum_i C_i O_i$. The coefficient $C_i$ is of the order 
of $\lam ^{-2}$ where
$\lam $ is the characteristic scale for the operator $O_i$. 
The operators $O_i$ of d=6 that can contribute to the process
$e^+e^-\rightarrow HZ$ are [2]

$$\eqalignno{O_1 &=(\bar{l}\gmmuu l){\imath\over 2}[\Phi^+(\dmuphi )-
(\dmuphi)^+\Phi ]\cr
&={v^2 g\over 4c_w}\bar{e}_L\gmmuu e_L
Z_{\mu}+{vg\over 2c_w}
\bar{e}_L\gmmuu e_L 
 Z_{\mu}H +... .&(1)\cr}$$

$$\eqalignno{O_2 &=(\bar{e}\gmmuu e){\imath\over 2}[\Phi^+(\dmuphi )-
(\dmuphi)^+\Phi ]\cr
&={v^2 g\over 4c_w}\bar{e}_R\gmmuu e_R Z_{\mu}+{vg\over 2c_w}
\bar{e}_R\gmmuu e_R Z_{\mu}H +... .&(2)\cr}$$

$$\eqalignno{O_3 &=(\bar{l}\gmmuu\taua l){\imath\over 2}[\Phi^+
\taua(\dmuphi )-
(\dmuphi)^+\taua\Phi ]\cr
&={v^2 g\over 4c_w}\bar{e}_L\gmmuu e_L
Z_{\mu}+{vg\over 2c_w}
\bar{e}_L\gmmuu e_L 
 Z_{\mu}H +... .&(3)\cr}$$

$$\eqalignno{O_4 &=(\bar {l}\dmu e)\dmuphi +h.c. ={\imath
 gv\over 2\sqrt {2}c_w}
[(\delmu \bar {e}_R)e_L-\bar {e}_L(\delmu e_R)]Z^{\mu}\cr
&+{\imath
 g\over 2\sqrt {2}c_w}
[(\delmu \bar {e}_R)e_L-\bar {e}_L(\delmu e_R)]HZ^{\mu}\cr
&+{\imath g s^2_w\over \sqrt {2}c_w}(\bar {e}_L e_R-\bar{e_R} e_L)
Z^{\mu}\delmu H +{\imath e\over \sqrt {2}}(\bar {e}_R e_L-\bar{e_L} e_R)
A^{\mu}\delmu H+... .&(4)\cr}$$

$$\eqalignno{O_5 &=(\dmu \bar{l}) e\dmuphi +h.c. ={\imath gv\over 2\sqrt {2}
c_w}
[\bar {e}_R(\delmu e_L)-(\delmu \bar {e}_L) e_R]Z^{\mu}\cr
&+{\imath g\over 2\sqrt {2}
c_w}
[\bar {e}_R(\delmu e_L)-(\delmu \bar {e}_L) e_R]HZ^{\mu}\cr
&+{\imath g(c^2_w-s^2_w)\over 2\sqrt {2}c_w}(\bar {e}_L e_R-\bar{e_R} e_L)
Z^{\mu}\delmu H +{\imath e\over \sqrt 2}(\bar {e}_L e_R-\bar{e_R} e_L)
A^{\mu}\delmu H+... .&(5)\cr}$$

$$\eqalignno{O_6 &=(\bar {l}\sigma^{\mu\nu}\tau_3 e)\phi W^3_{\mu\nu}+h.c.\cr
&=-{1\over \sqrt {2}}(\bar {e}_L\sigma^{\mu\nu}e_R)H(s_w F_{\mu\nu}
+c_w Z_{\mu\nu})+ h.c. +... .&(6)\cr}$$

$$\eqalignno{O_7 &=(\bar {l}\sigma^{\mu\nu} e)\phi B_{\mu\nu}+h.c.\cr
&={1\over \sqrt 2}(\bar {e}_L\sigma^{\mu\nu}e_R)H(c_w F_{\mu\nu}
-s_w Z_{\mu\nu})+ h.c. +... .&(7)\cr}$$

In the above we have expressed $\phi$ in unitary gauge and have written
only the resulting
 d=4, 5 and 6 operators involving charged
lepton, scalar and neutral gauge fields that can contribute to precision
measurements on Z peak and to $e^+e^-\rightarrow HZ$. On transforming 
the lepton fields from the 
gauge eigenstate basis to the mass eigenstate basis
the operators $O_6$ \& $O_7$ give rise to  d=5 FCNC operators involving
both $Z_{\mu}$ and $A_{\mu}$. In particular the coefficients $C_6$
and $C_7$ must satisfy the strong experimental bound [4] on the branching
fraction for the process $\mu\rightarrow e\gamma $ which implies that

$${\Gamma_{\mu\rightarrow e\gamma}\over \Gamma_{\mu\rightarrow e\bar{\nu}_e
 \nu_{\mu}}}\approx {6\pi^2 v^6\over \lam ^4m_{\mu}^2}(c_w-s_w)^2
\sin ^2 \theta_{12}\le 5\times 10^{-11}.\eqno(8)$$

If we assume that $\sin\theta_{12}\approx .2$, the scale $\lam\approx
C_6^{-{1\over 2}}\approx C_7^{-{1\over 2}}$ associated with $O_6$ and 
$O_7$ must be greater than 3500 Tev. Such a huge scale for $O_6$ and 
$O_7$ can be avoided by assuming a symmetry that forbids FCNC vertices
upon transformation from the gauge basis
to the mass basis. However the flavor diagonal
terms of $O_6$ and $O_7$ must still satisfy the constraint [5] arising
from the experimental value of ${g_e-2\over 2}$ which implies that

$$[({\delta g_e\over 2})_{expt}-({\delta g_e\over 2})_{sm}]=
({\delta g_e\over 2})_{new}\approx {2\sqrt {2}m_e v(c_w-s_w)\over
\lam ^2 e}\le .27\times 10^{-9}.\eqno(9)$$

Hence the scale $\lam $ associated with the flavor diagonal terms of
$O_6$ and $O_7$ must be greater than or of the order of 40 Tev which is
considerably greater than the bound (2.5 Tev) on the scale associated with
$O_1$, $O_2$ and $O_3$ that follows from Z pole precision measurements.
We shall therefore ignore effect of $O_6$ \& $O_7$
 on the process $e^+e^-\rightarrow HZ$. The remaining operators $O_i$, i=1-5
do not contain any ee$\gm$ vertex but they do contain eeZ vertex. The scale
associated with these flavor diagonal operaors are best constrained
by precision measurements on Z pole. However $O_1$, $O_2$ and $O_3$
affect Z pole physics through d=4 operators, but $O_4$ \& $O_5$
contribute to the same through d=5 operators. Hence the constraint
on $C_4$ and $C_5$ that follows from Z pole precision measurements
is expected to be weaker than that on $C_1$, $C_2$ and $C_3$.
If we assume that $C_4=C_5={1\over \lam ^2}$  we get

$$C_4 O_4 + C_5 O_5\approx {\imath C_4g\over \sqrt {2} c_w}(\bar {e}
\gamma_5 e)Z^{\mu}\delmu H.\eqno(10)$$

where we have integrated by parts and have used the relation
$\delmu Z^{\mu}=0$ for on shell Z boson. 
Note that $O_4$ and $O_5$ contributes to Z pole precision measurements
through d=5 terms in contrast to $O_1-O_3$ which contributes to the same
via d=4 terms. Hence the constraint on $C_4$ and $C_5$
 that follows from Z pole precision measurements is expected
to be slightly weaker than that  on $C_1$, $C_2$ and $C_3$.
In this work we shall determine the scale associated with $O_1$,
$O_2$ and $O_3$ from the LEP and SLD data and equate it to the
scale associated with $O_4$ and $O_5$.
The effective Lagrangian that contributes to the process
 $e^+e^-\rightarrow HZ$ becomes

$$\eqalignno{L_{eff}&={g\over 2c_w}[(1-2s_w^2)+{v^2\over 2}(C_1+C_3)]
\bar {e}_L\gmmuu e_L Z_{\mu}
+{g\over 2c_w}[{v^2\over 2}C_2-2s_w^2] 
\bar {e}_R\gmmuu e_R Z_{\mu}\cr
&+{v g^2\over 4 c_w^2}HZ_{\mu}Z^{\mu}
+(C_1+C_3){vg\over 2c_w}\bar {e}_L\gmmuu e_L Z_{\mu}\cr
&+C_2{vg\over 2c_w} \bar {e}_R\gmmuu e_R Z_{\mu}+{\imath C_4 g\over
 \sqrt {2}c_w}\bar {e}\gamma_5 e Z^{\mu}\delmu H.&(11)\cr}$$

To a first approximation we can neglect the small corrections (of the order
of 1\%)to the $Ze\bar {e}$ vertex due to new physics.
The total cross-section for $e^+e^-\rightarrow HZ$ is given by
$$\eqalignno{\sigma_T &= \sigma_{LR+RL}+\sigma_{LL+RR}\cr
&={1\over 384\pi s}{f(s,M_z,M_H)\over s}[\{{2\pi\alp\over s_w^2c_w^2}
{(1-2s_w^2)\over (s-M_z^2)}+(C_1+C_3)\}^2\cr
&+\{C_2-{2\pi\alp\over s_w^2c_w^2}\}^2][12sM_z^2+f^2(s,M_z,M_H)]\cr
&+{1\over 64\pi v^2}{f(s,M_z,M_H)\over s}C_4^2 f^2(s,M_z,M_H).&(12)\cr}$$

where $f^2(s,M_z,M_H)=(s+M_z^2-M_H^2)^2-4sM_z^2$. The first term arises
from $e^-_Le^+_R+e^-_Re^+_L\rightarrow HZ$ and the second term
from $e^-_Le^+_L+e^-_Re^+_R\rightarrow HZ$.
We shall consider two distinct cases of new physics effects on $\sigma_T$.
In the first scenario we shall assume that new physics effects on
$\sigma_{LR+RL}$ interferes constructively with those of SM. A typical
example of this scenario is $C_1=C_3=C_4=-{C_2\over 2}={1\over \lam ^2}$.
In the second scenario the new physics effects on $\sigma_{LR+RL}$
will be assumed to interfere destructively with SM effects, an example of
 which is $C_1=C_3=-C_4=-{C_2\over 2}=-{1\over \lam ^2}$. We shall assume
a common value of $\lam $=2.5 Tev for both scenarios.

The operators $O_1$, $O_2$ and $O_3$ give rise to small corrections to
 the vector and axial vector couplings of Z boson to $e \bar {e}$. It
is therefore important to consider the constraints on $C_1$, $C_2$
and $C_3$ that follow from LEP and SLD asymmetry measurements on Z peak
and whether these constraints are consistent with the value of 
$\lam $ assumed above. We have $g^e_v=(g^e_v)_{sm}-{v^2\over 8}(C_1+C_2
+C_3)$ and $g^e_a=(g^e_a)_{sm}+{v^2\over 8}(C_1+C_3-C_2)$. Hence in both
scenarios $\delta g^e_v\approx 0$ and $\delta g^e_a \approx -{v^2\over 4}C_2$
i.e. new physics effects renormalizes the weak axial charge of the electron
but does not affect its weak vector charge. On Z peak [6]
we have
 $\alr \approx (\alr )_{sm}[1-{\delta g^e_a\over (g^e_a)_{sm}}]$ and
 $\aefb \approx
(\aefb )_{sm}[1-2{\delta g^e_a\over (g^e_a)_{sm}}]$ provided
$\vert {\delta g^e_v\over (g^e_v)_{sm}}\vert \ll
\vert {\delta g^e_a\over (g^e_a)_{sm}}\vert $.
It then follows that ${(\delta \alr )_{new}\over (\alr )_{sm}}\approx
\pm {v^2\over 2\lam^2 (g^e_a)_{sm}}$ and 
 ${(\delta \aefb )_{new}\over (\aefb )_{sm}}\approx\pm
 {v^2\over \lam^2 (g^e_a)_{sm}}$ where the upper (lower) sign corresponds
to destructive (constructive) interference scenario.
 The average values of $\alr $
and $\aefb $ reported by SLD and LEP [7] are $\alr =.156\pm .008$
and $\aefb =.0160\pm .0024$. The predicted values for these asymmetries
in the context
 of the SM are
$(\alr )_{sm}=.142\pm .003\pm.003$ and $(\aefb )_{sm}=.0151$.
The experimental bounds on $\dalr $ and $\daefb $ are ${\dalr \over 
(\alr )_{sm}}\approx .099$ and 
${\daefb \over (\aefb )_{sm}}\approx .06$. For $\lam \approx 2.5$ Tev 
the new physics contributions are given by 
 ${\dalr \over (\alr )_{sm}}\approx \pm .02$ and
${\daefb \over (\aefb )_{sm}}\approx \pm .04$ both of which are clearly
compatible with the experimental bounds. On the contrary the contribution
of $O_4$ to $\alr $ at the Z pole is given by ${\dalr\over (\alr )_{sm}}\approx
-{1\over 2}{({g\over 4c_w})^2 {v^2s\over \lam^4}\over (\gel )^2 +(\ger )^2}$.
The current SLD precision for measuring $\alr $ is about 5\%. Hence even if
$\lam $ is as low as 400 Gev the contribution of $O_4$ to $\alr $
is far too small (about .78\%) to be detected at SLD. In the following
 we shall equate the scale of $O_4$ and $O_5$ to that of
 $O_1$, $O_2$ and $O_3$.
 This will give us a lower bound on the cross-section for the process
$e^+e^-\rightarrow HZ$.
 Further if the positive sign of
the phenomenological bounds on $\dalr $ and $\daefb $ is taken seriously,
experiments would seem to prefer the destructive interference scenario
considered in this article,
implying that the observed higgs production cross-section will be lower
than that of the SM.

From eqn. 11 we find that for $\sqrt {s} =500 $ Gev and $m_H =
150$ Gev, $\sigma_{T}\approx 160 $ fb (43 fb) for the constructive 
(destructive) interference case. This is to be compared with the
SM prediction of $\sigma_{sm}\approx 52$ fb. When $m_H$ is increased
to 200 Gev, $\sigma_T$ drops to 128 fb in the constructive case and to
33 fb in the destructive case. The corresponding value of the
cross-section in the context of SM is 42 fb. In the case of constructive
interference the dominant contribution to $\sigma_T$ 
 comes from $\sigma_{LR+RL}$ whereas
for destructive interference $\sigma_{LL+RR}$ forms the dominant part.
$\sigma_T$ is therefore quite insensitive to the value of $C_4$ in the
former case but depends quite strongly on it in the latter.
 For an integrated luminosity
of 30 fb $^{-1}$  [8] the effect of new physics would be to increase 
the number of HZ events
 by 3060 for constructive interference or decrease it by 390 for 
destructive interference. 
Note that at low energies ($\sqrt {s}\approx 500 $ Gev)
the dominant production mechanism for intermediate mass higgs boson 
at an $e^+e^-$ collider is
$e^+e^-\rightarrow HZ$ [8]. For higher energies ($\sqrt {s}\approx 1$ Tev)
the dominant production mechanism becomes $e^+e^-\rightarrow \nu_e
\bar {\nu}_e H$. However
new physics affects the latter process only through small
corrections (of the order of 1\%) to the usual SM vertices. There is no 
  d=5 or 6  operator for the vertex 
$\bar {e}e\bar {\nu}_e \nu_e H$ similar to
the $\bar {e}eZH$ vertex. Hence the overall effect of Tev scale
new physics
on the process  $e^+e^-\rightarrow \nu_e\bar {\nu}_e H$ is expected to be
much smaller than that on $e^+e^-\rightarrow HZ$.

Some comments are in order about the reliability of the
 value of $\lam $, the scale of
 new physics, used in our analysis. The scale $\lam $ for new
interactions should be related [9]
 to the amplitude v (v plays the role of $f_{\pi}$ in electro-weak
 theory) for producing scalar
particles out of the vacuum through the relation ${\lam\over v}=g_s$
where $g_s$ is the induced coupling for the low energy theory.
According to theoretical estimates $g_s$ is expected to lie
between 1 and $4\pi $. In low energy 
 QCD for example $g_{\rho}={M_{\rho}\over 
f_{\pi}}\approx 6 $.
 For $\lam \approx 2.5$ Tev we find that $g_s\approx 10$ which is in 
agreement  with our theoretical expectations about new interactions 
underlying the EW theory.

\centerline{\bf References}

\item{1.} The quadratic instability of the higgs sector was first
pointed out by K. Wilson, unpublished.
 E. Gildener, Phys. Rev. D 14, 1667 (1976); L. Susskind,
Phys. Rev. D 20, 2619 (1979).

\item{2.} W. Buchmuller and D. Wyler, Nucl. Phys. B 268, 621 (1986).

\item{3.} J. Gunion, S. Dawson, H. Haber and G. Kane, ``The Higgs
Hunters Guide'', Addison Wesley (Menlo-Park, 1991).

\item{4.} Review of Particle Properties, Phys. Rev. D 54, 21 (1996).

\item{5.} T. Kinoshita, Phys. Rev. Lett. 47, 1573 (1981); Review
of Particle Properties, Phys. Rev. D 54, 21 (1996).

\item{6.} V. Barger and R. Philips, ``Collider Physics'', Addision-Wesley
(Menlo-Park, 1987).

\item{7.} Review of Particle Properties, Phys. Rev. D 54, 222 (1996).

\item{8.} Junichi Kanzaki, in Proceedings of the second workshop on 
JLC, ed. by S. Kawabata (KEK 91-10).

\item{9.} H. Georgi, Phys. Lett. B 298, 187 (1993); R. S. Chivukula
and E. H. Simmons, Phys. Lett. B 388, 788 (1996).

\vfill\eject
\centerline{\bf Figure Captions}

\item{Fig. 1.} The cross-section for the process $e^+e^-\rightarrow HZ$
for $m_H=150$ Gev plotted against $\sqrt {s}$. (a) constructive
interference (b) standard model and (c) destructive interference.

\item{Fig. 2} Same as Fig. 1 with $m_H=200$ Gev.
 
\end